\documentstyle[12pt,fleqn,epsfig]{article}
\newcommand{\beq}{\begin{equation}}
\newcommand{\eeq}{\end{equation}}

\def\tE{\tilde{E}}

\def \ut#1{\rlap{\lower1ex\hbox{$\sim$}}#1{}}
\def \UT#1{\rlap{\lower1ex\hbox{\scriptsize$\sim$}}#1{}}
\newcommand{\UI}[1]{^{\mbox{ } #1}}

\def\ep{\epsilon}

\def\utep{\UT{\epsilon}}

\def\Tr{{\rm Tr}}
\begin{document}
\begin{flushright}
OU-HET/223\\gr-qc/9510019\\October 1995
\end{flushright}
\vspace{0.5in}
\begin{center}\Large{\bf Multi-plaquette solutions for\\
discretized Ashtekar gravity}\\
\vspace{1cm}\renewcommand{\thefootnote}{\fnsymbol{footnote}}
\normalsize\ Kiyoshi Ezawa\footnote[1]
{Supported by JSPS.
e-mail address: ezawa@funpth.phys.sci.osaka-u.ac.jp}
        \setcounter{footnote}{0}
\vspace{0.5in}

        Department of Physics \\
        Osaka University, Toyonaka, Osaka 560, Japan\\
\vspace{0.1in}
\end{center}
\vspace{1.2in}
\baselineskip 17pt
\begin{abstract}

A discretized version of canonical quantum gravity proposed by
Loll is investigated.
After slightly modifying Loll's discretized Hamiltonian constraint,
we encode its action on the spin network states in terms of
combinatorial topological
manipulations of the lattice loops.
Using this topological formulation we find new solutions
to the discretized Wheeler-Dewitt equation.
These solutions have their
support on the connected set of plaquettes.
We also show that these solutions are not
normalizable with respect to the induced heat-kernel measure
on $SL(2,{\bf C})$ gauge theories.

\end{abstract}
\newpage
\baselineskip 20pt

As an approach to quantize gravity nonperturbatively, canonical
quantization of general relativity  has been investigated
for more than thirty years.
In the conventional metric formulation\cite{ADM},
we have not yet found any solutions for its basic equation,
i.e. the Wheeler-Dewitt (WD) equation\cite{dewitt}
(or the Hamiltonian constraint).
This had been one of the serious
obstructions against smooth progress
in this approach for a long time.
The situation was drastically changed after the discovery of
Ashtekar's new canonical variables in 1986\cite{ashte}.
Ashtekar's variables consist of the complex-valued $SU(2)$ connection
$A_{a}^{i}$ and the densitized triad $\tE^{ia}$.
Using these variables the Hamiltonian
constraint takes the simple form:
\beq
{\cal H}=\ep^{ijk}F_{ab}^{i}\tE^{ja}\tE^{kb},\label{eq:hami}
\eeq
where $F_{ab}^{i}$ is the curvature of the connection $A_{a}^{i}$.
The WD equation in terms of
new variables are therefore expected to
have solutions. Indeed several
types of solutions has been constructed
using Wilson loops defined on
smooth loops with or without intersections
\cite{jacob}\cite{H}\cite{BP}\cite{ezawa}.

These solutions are, however, of little interest.
Because they are already eliminated by the operator
$\ep^{ijk}\utep_{abc}\tE^{ja}\tE^{kb}$,
naively we consider that they
correspond to the states with degenerate metric.
This suggests that
we have to search for the solutions
on which the action of the curvature
plays an essential role.
In terms of Wilson loops, multiplication
by the curvature is encoded by
the action of the area derivative\cite{migdal}.
It is therefore important to define the area derivative
without any ambiguity.
This seems to be a nontrivial problem in the continuum approach.

In the lattice formulation, we can in principle express
the area derivative by the operation of inserting a plaquette
to the lattice loop arguments. The lattice formulation therefore
deserves studying as a heuristic model of the continuum approach.

A discretized version of Ashtekar's formalism was proposed by
Loll\cite{loll}. This model is defined on a 3 dimensional
cubic lattice of size N. We will follow the notations
in ref.\cite{loll} and label lattice sites by $n$ and
three positive directions of links by $\hat{a}$.
The connection $A_{a}^{i}$ is replaced by
the link variables $V(n,\hat{a})$ which takes the value in
$SL(2,{\bf C})$ and the conjugate momenta $\tE^{ia}$ is replaced
by the left translation operator $p_{i}(n,\hat{a})$.
Their commutation relations are:
\begin{eqnarray*}
[V(n,\hat{a}),V(m,\hat{b})]=0,\quad
[p_{i}(n,\hat{a}),V(m,\hat{b})]=-\frac{i}{2}
\delta_{n,m}\delta_{\hat{a}\hat{b}}(\tau_{i}V(m,\hat{b})),\\
{}[ p_{i}(n,\hat{a}),p_{j}(m,\hat{b}) ] =i
\delta_{n,m}\delta_{\hat{a}\hat{b}}\ep_{ijk}
p_{k}(n,\hat{b}),
\end{eqnarray*}
where $\tau_{i}$ equals to $-i$ times of the Pauli matrices.

Among the three constraints in Ashtekar's formalism\cite{ashte},
the Gauss' law constraint is solved
by considering only the gauge-invariant
functionals of link variables, namely, spin network states
\cite{penrose}\cite{baez}\cite{RS}.
The diffeomprhism constraint is formally solved by regarding our
lattice to be a purely topological object\cite{loll}
(see also \cite{lewa}). Thus we are left only with the Hamiltonian
constraint (\ref{eq:hami}), whose discretized form proposed by
Loll is:
\beq
H^{C}(n)=\sum_{\hat{a}<\hat{b}}\ep^{ijk}p_{i}(n,\hat{a})
p_{j}(n,\hat{b}){\rm Tr}(V(n,P_{\hat{a}\hat{b}})\tau_{k}),
\label{eq:loll}
\eeq
where $V(n,P_{\hat{a}\hat{b}})\equiv V(n,\hat{a})V(n+\hat{a},\hat{b})
V(n+\hat{b},\hat{a})^{-1}V(N,\hat{b})^{-1}$ denotes a plaquette loop.

This definition obviously lacks symmetry; only positive
directions emanating from the site $n$ are subject to the action
of $H^{C}(n)$. This is not desirable because there are privileged
directions in the world.

In order to provide symmetric forms
of discretized Hamiltonian constraint,
we first introduce link variables in the negative direction
$V(n,-\hat{a})\equiv V(n-\hat{a},\hat{a})^{-1}$ and the right
translation operator $p_{i}(n,-\hat{a})$:
\beq
p_{i}(n,-\hat{a})V(n,-\hat{a})=-\frac{i}{2}\tau_{i}V(n,-\hat{a}),
\quad
p_{i}(n,-\hat{a})V(n-\hat{a},\hat{a})=
\frac{i}{2}V(n-\hat{a},\hat{a})\tau_{i}.
\eeq
Naively considering, we think of two candidates for the symmetric
discretized Hamiltonian constraint. One is
\beq
H_{I}^{C}(n)=\sum_{\hat{a}<\hat{b}}\ep^{ijk}
\Tr(\widetilde{V}(n,\hat{a}\hat{b})\tau_{k})
(p_{i}(n,\hat{a})-p_{i}(n,-\hat{a}))
(p_{j}(n,\hat{b})-p_{j}(n,-\hat{b})),\label{eq:hamiI}
\eeq
where $\widetilde{V}(n,\hat{a}\hat{b})\equiv\frac{1}{4}
(V(n,P_{\hat{a}\hat{b}})+V(n,P_{\hat{b},-\hat{a}})+
V(n,P_{-\hat{a},-\hat{b}})+V(n,P_{-\hat{b},\hat{a}}))$.
The other is
\beq
H_{II}^{C}(n)=\sum_{\hat{a}<\hat{b}}\sum_{\eta_{1},\eta_{2}=\pm}
\ep^{ijk}\Tr(V(n,P_{\eta_{1}\hat{a},\eta_{2}\hat{b}}))
p_{i}(n,\eta_{1}\hat{a})p_{j}(n,\eta_{2}\hat{b}).\label{eq:hamiII}
\eeq

Probably $H_{I}^{C}$ is more preferable than $H_{II}^{C}$ because
the area derivative in the former is uniformly expressed by
the insertion of $\frac{1}{2}(\widetilde{V}-\widetilde{V^{-1}})$.
Indeed the result of the action of $H_{I}^{C}$ is identical,
up to the overall factor, to
the action of the continuum Hamiltonian under the regularization
used in ref.\cite{ezawa}. The WD equation
using $H_{I}^{C}$ therefore has solutions which are the lattice
analog of the solutions found in
refs.\cite{jacob}\cite{H}\cite{BP}\cite{ezawa},
provided that the smooth loops are
replaced by straight Polyakov loops.

Our purpose is, however, to find out \lq\lq nontrivial solutions"
which becomes the solution only after
the area derivative is taken into account.
To this end it is much easier to use
$H_{II}^{C}$, because the terms appearing in $H_{I}^{C}$ is naively
four times as many as those in $H_{II}^{C}$.
While $H_{II}^{C}$ may bb
less suitable to regularize the continuum theory, we expect that
it will provide some essential lessons
concerning to the nontrivial solutions.
For these reasons we will henceforth deal only with $H_{II}^{C}$.

The action of the discretized Hamiltonian
constraint can be computed
by using only the $SL(2,{\bf C})$ algebra. The two
identities are particularly useful:
\beq
\tau_{i}\tau_{j}=-\delta_{ij}+\ep_{ijk}\tau_{k},\quad
(\tau_{i})_{A}\UI{B}(\tau_{i})_{C}\UI{D}=
\delta_{A}\UI{B}\delta_{C}\UI{D}-2\delta_{A}\UI{D}\delta_{C}\UI{B}.
\nonumber
\eeq
As in the continuum case, the nonvanishing contributions are
obtained only from the series of link variables
with kinks or intersections.
Because the Hamiltonian constraint involves second order derivative,
it is convenient to separate its action as follows:
\beq
H_{II}^{C}(n)=H_{II}^{C}(n)_{1}+H_{II}^{C}(n)_{2},
\eeq
where $H_{II}^{C}(n)_{1}$ is the
sum of the action on the single series and
$H_{II}^{C}(n)_{2}$ is the sum of the action on the pairs of series.
The problem of evaluating the action of the discretized Hamiltonian
constraint $H_{II}^{C}$ is thus reduced to
that of calculating the action on all possible
types of kinks and intersections involving at most two
series of links.

For example, the action on kinks is as follows
($\hat{a}\neq\hat{b}$):
\begin{eqnarray}
H_{II}^{C}(n)_{1}\cdot V(n-\hat{a},\hat{a})V(n,\hat{b})
&\!\!\!=\!\!\!&
-\frac{1}{2}V(n-\hat{a},\hat{a})(V(n,P_{-\hat{a},\hat{b}})
-V(n,P_{\hat{b},-\hat{a}}))V(n,\hat{b}),\nonumber \\
H_{II}^{C}(n)_{2}\cdot\left(
V(n-\hat{a},\hat{a})V(n,\hat{b})\right.
&\!\!\!\otimes\!\!\!&\left.
V(n-\hat{a},\hat{a})V(n,\hat{b})\right)=0.\label{eq:topo}
\end{eqnarray}
Topologically, the former action can be interpreted as taking the
difference of two operations of inserting plaquettes with the
opposite orientations.
The action on the other types of vertices can also be interpreted in
terms of combinatorial topology. We will depict the action on some
typical vertices in figure1, where the bold lines stand for
the series of link variables.
Once we express the action of $H_{II}^{C}$
in terms of combinatorial topology, the orientation of the
curve is irrelevant because of the symmetry of $H_{II}^{C}$.

\begin{figure}[t]
\begin{center}
\epsfig{file=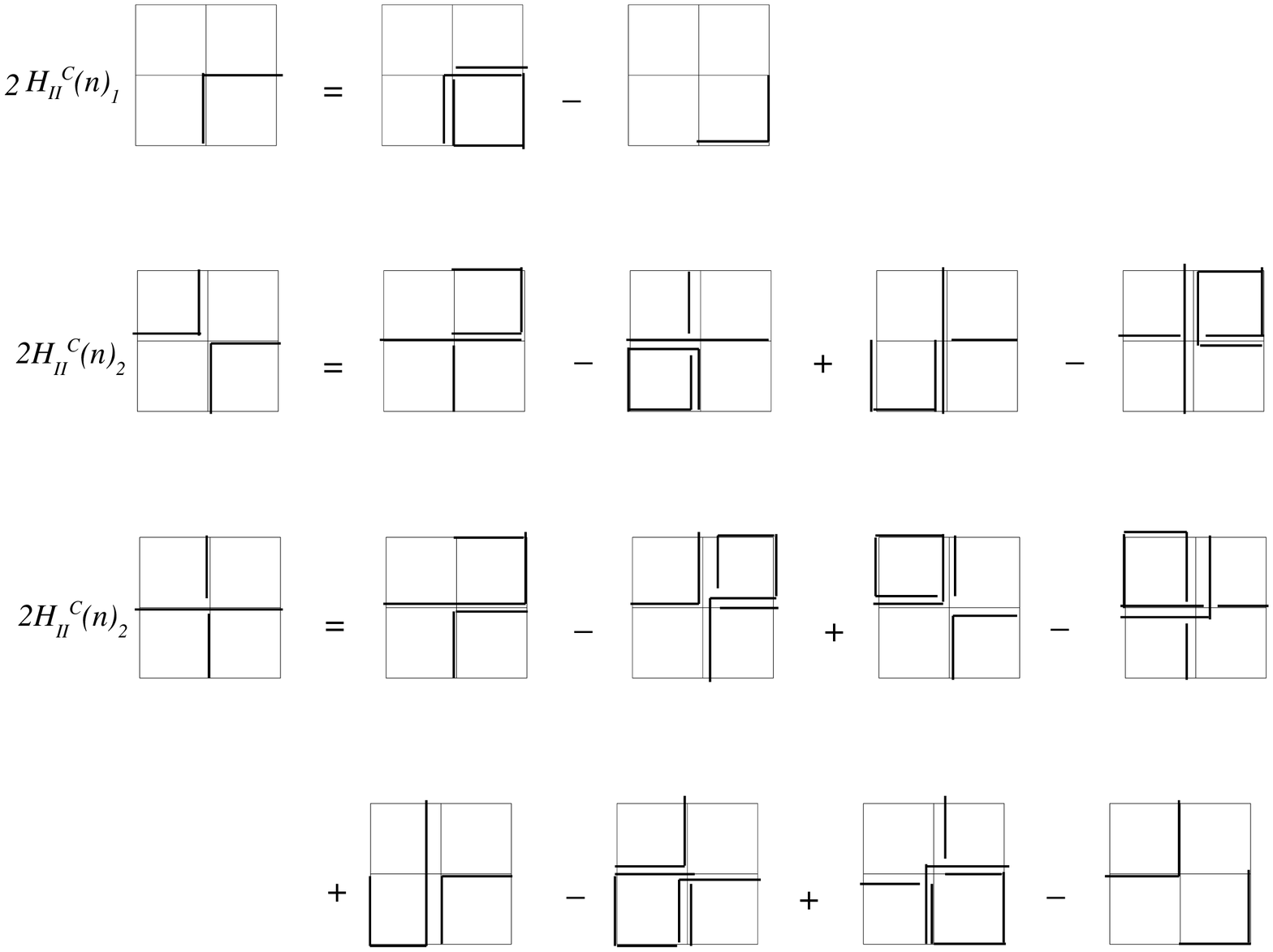,height=12cm}
\end{center}
\caption{Action of $H^{C}_{II}$ on some typical vertices}
\end{figure}

A merit of this topological formulation is that we can visualize
the action of the Hamiltonian constraint. It is expected that
we can fully exploit this merit in finding
solutions to the WD equation\footnote{
A weakness of the topological formulation is that we have to
work with the overcomplete basis of wavefunctions. In this respect,
it would be better to describe
the action of the Hamiltonian constraint
in terms solely of spin network states, which are known to
form a complete basis\cite{baez}\cite{FK}.
In the spin network description,
however, we have to deal with tedious linear combinations consisting
of considerably many spin network states. This description is not
considered to be suitable for the visual search of solutions, while
it may be useful for computer analysis.}.

As an exercise we will provide a set of the simplest
\lq\lq nontrivial solutions"
on which the contribution of area derivative is
essential.
We first consider the action on the trace of the $l$-th power of the
plaquette $\Tr( V(n,P_{\hat{a}\hat{b}})^{l})$.
This is calculated by using
eq.(\ref{eq:topo})
\beq
H_{II}^{C}(n)\cdot\Tr( V(n,P_{\hat{a}\hat{b}})^{l})=
\frac{l}{2}\left(\Tr( V(n,P_{\hat{a}\hat{b}})^{l+1})
-\Tr( V(n,P_{\hat{a}\hat{b}})^{l-1})\right).\label{eq:power}
\eeq
This seems to imply the following equation
\beq
H_{II}^{C}(n)\cdot\sum_{k=1}^{\infty}\frac{-2}{2k+1}
\Tr( V(n,P_{\hat{a}\hat{b}})^{2k+1})=2.\label{eq:series}
\eeq
While the issue of the convergence remains, this can be resolved
by exploiting the idea of analytic continuation. More explicitly
we consider as follows. First we reinterpret eq.(\ref{eq:power}) as
\beq
H_{II}^{C}(n)\cdot\Tr F(V(n,P_{\hat{a}\hat{b}}))=\Tr
\left[\frac{V(n,P_{\hat{a}\hat{b}})^{2}-1}{2}
\frac{d}{dV}F(V(n,P_{\hat{a}\hat{b}}))\right],
\eeq
where $F$ denotes an arbitrary polynomial.
We can readily extend this equation
to the case where $F$ is a function which can be expressed by
a Laurent series. Thus we find
\beq
H_{II}^{C}(n)\cdot\Tr\log\left(\frac{1-V(n,P_{\hat{a}\hat{b}})}
{1+V(n,P_{\hat{a}\hat{b}})}\right)=2.\label{eq:log}
\eeq
The power series expansion of the expression on the l.h.s yields
the l.h.s of eq.(\ref{eq:series}).
As for the action of $H_{II}^{C}(m)$
with $m\neq n$, the following can be said. When $m$ coincides with
one of the vertices of the plaquette $P_{\hat{a}\hat{b}}$,
the result is identical to eq.(\ref{eq:log}) owing to the symmetry
of $H_{II}^{C}$. When $m$ does not coincide, on the other hand, the
action necessarily vanishes.
Putting these results together, we find
\beq
H_{II}^{C}(m)\cdot\Tr\log\left(\frac{1-V(n,P_{\hat{a}\hat{b}})}
{1+V(n,P_{\hat{a}\hat{b}})}\right)=\left\{
\begin{array}{cc}2&
\mbox{for $m=n,n+\hat{a},n+\hat{b},n+\hat{a}+\hat{b}$,}\\
0&\mbox{for $m\neq n,n+\hat{a},n+\hat{b},n+\hat{a}+\hat{b}$.}
\end{array}\right. \label{eq:master}
\eeq
Now we can provide the prescription for constructing
\lq\lq multi-plaquette solutions" on which
the action of the area derivative is essential:
i) prepare a connected set of plaquettes $\{P\}$in which each vertex
belongs to at least two plaquettes; ii) assign to
each plaquette $P$ a weight factor $w(P)$ so that the sum of
weight factors of the plaquettes which meet at each vertex vanishes;
iii) the following expression yields a solution
\beq
<A|\{w(P)\}>\equiv
\sum_{P\in\{P\}}w(P)\Tr\log\left(\frac{1-V(P)}{1+V(P)}\right).
\label{eq:mpla}
\eeq
The two simplest assignments of the weights $\{w(P)\}$ are
shown in figure 2. Because the product
$<A|\{w(P)\}><A|\{w(P^{\prime})\}>$ is also a solution if
the sets $\{P\}$ and $\{P^{\prime}\}$ are disconnected, we found
a considerably large number of
solutions to the discretized WD equation
(\ref{eq:hamiII})\footnote{
We can construct a set of solutions to (\ref{eq:loll}) by
a similar procedure.
However, the analogous prescription cannot apply to
eq.(\ref{eq:hamiI}). From this we expect that the continuum limit
of $<A|\{w(P)\}>$ are not solutions to the continuum WD equation,
at least in a naive regularization.}.

\begin{figure}[t]
\begin{center}
\epsfig{file=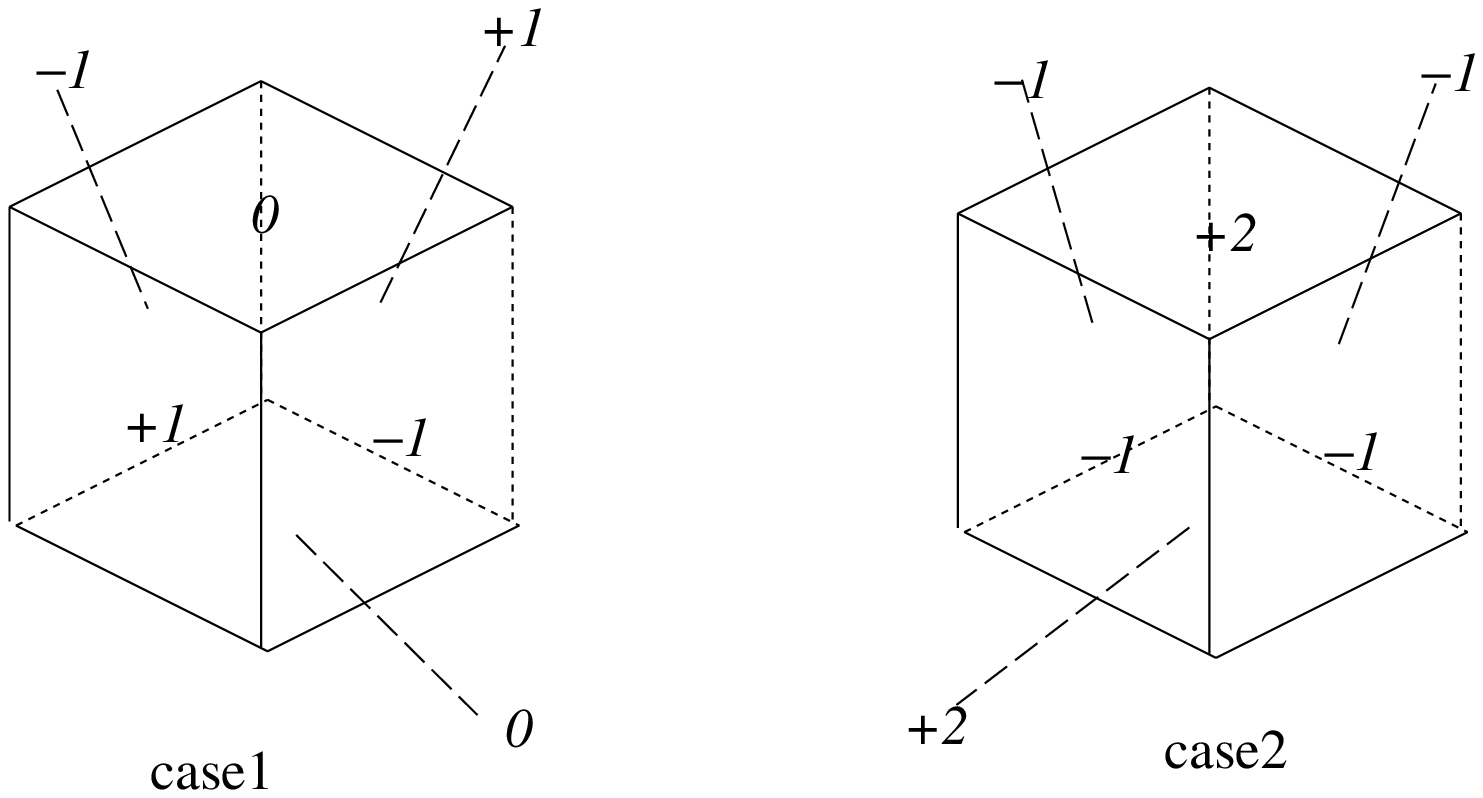,height=6cm}
\end{center}
\caption{Two simplest assignments of weight factors.}
\end{figure}

One may wonder whether or not these solutions are normalizable
with respect to an appropriate inner product. To investigate this
problem, it is convenient to translate the result into spin network
states. In the present case, they are nothing but symmetrized traces
$$
\Tr S(V(P)^{k})\equiv
V(P)_{(A_{1}}\UI{B_{1}}\cdots V(P)_{A_{k})}\UI{B_{k}}.
$$
The action of $H_{II}^{C}$ on these states are calculated as
$$
H_{II}^{C}(n)\Tr S(V(n,P_{\hat{a}\hat{b}})^{k})=
\frac{k}{2}S(V(n,P_{\hat{a}\hat{b}})^{k+1})
-\frac{k+2}{2}S(V(n,P_{\hat{a}\hat{b}})^{k-1})
$$
Thus, to obtain multi-plaquette solutions $<A|\{w(P)\}>_{S}$
in the spin network representation,we have only to replace
$\log(\frac{1-V}{1+V})$ in eq.(\ref{eq:mpla}) by
\beq
<A|P>_{S}\equiv
\sum_{m=0}^{\infty}\frac{1}{(2m+3)(2m+1)}\Tr S(V(P)^{2m+1}).
\label{eq:1pla}
\eeq

Let us now investigate the
normalizability of multi-plaquette solutions.
We first look into the induced Haar measure\cite{baez}\cite{asle}
regarding our model as a sort of $SU(2)$ gauge theory.
Owing to the consistency property of the induced Haar measure
and the orthogonality of
the spin network states with different numbers of link variables,
we have only to investigate the norm of a
one-plaquette state (\ref{eq:1pla})(which is not a solution)
$$
\|<A|P>_{S}\|_{H}=\int d\mu_{H}(V(P))|<A|P>_{S}|^{2},
$$
where $d\mu_{H}$ is the Haar measure on $SU(2)$. We estimate this
norm as follows. Using bi-$SU(2)$ invariance of $d\mu_{H}$:
$d\mu_{H}(gVh)=d\mu_{H}(V)$ with $V,g,h\in SU(2)$,
and the fundamental identity
$\ep^{AC}V_{A}\UI{B}V_{C}\UI{D}=\ep^{BD}$,
the integration by the Haar measure is determined uniquely
\beq
\int d\mu_{H}(V)\prod_{i=1}^{2n}V_{A_{i}}\UI{B_{i}}=
\frac{1}{2^{n}(n+1)!n!}\sum_{\sigma\in P_{2n}}\prod_{k=1}^{n}
\ep_{A_{\sigma_{2k-1}}A_{\sigma_{2k}}}
\ep^{B_{\sigma_{2k-1}}B_{\sigma_{2k}}},
\eeq
where $P_{2n}$ is the group of permutations of $2n$ entries.
The integration of $\prod_{i=1}^{2n+1}V_{A_{i}}\UI{B_{i}}$
vanishes identically. From this equation we find
\beq
\int d\mu_{H}(V)\overline{\Tr S(V^{n})}\Tr S(V^{m})=\delta_{nm},
\label{eq:Haar}
\eeq
where the bar denotes the complex conjugate.
The desired norm is calculated from eqs.(\ref{eq:1pla})
(\ref{eq:Haar}). The result is
\beq
\|<A|P>_{S}\|_{H}=
\sum_{m=0}^{\infty}\frac{1}{(2m+3)^{2}(2m+1)^{2}}<\infty,
\eeq
i.e. the multi-plaquette solutions are normalizable w.r.t. the
induced Haar measure.

Next we look into the induced heat-kernel measure
$d\nu_{t}$\cite{ash2}\cite{hall} on $SL(2,{\bf C})$ gauge theories.
Because $d\nu_{t}$ is also bi-$SU(2)$
invariant and possesses the consistency property, we can
show that the spin network states with different numbers of
link variables are orthonormal w.r.t. $d\nu_{t}$, in particular
\beq
\int d\nu_{t}(V)
\overline{\Tr S(V^{n})}\Tr S(V^{m})=C(n)\delta_{nm}.
\eeq
There seems to be no algebraic principle which determines the
constant factor $C(n)$. Using some facts on the coherent-state
transform $C_{t}:L^{2}(SU(2),d\mu_{H})\rightarrow
L^{2}(SL(2,{\bf C}),d\nu_{t})^{\cal H}$
(Theorem 2 and eq.(30) of ref.\cite{hall}), however,
we can exactly estimate the constant factor
\beq
C(n)= e^{\frac{n(n+2)}{4}t}.
\eeq
As a result, the
one-plaquette states are not normalizable w.r.t. the
induced heat-kernel measure $d\nu_{t}$:
\beq
\int d\nu_{t}(V(P))|<A|P>_{S}|^{2}=
\sum_{m=0}^{\infty}
\frac{e^{\frac{n(n+2)}{4}t}}{(2m+3)^{2}(2m+1)^{2}}
\rightarrow \infty.
\eeq

We should note that the multi-plaquette solutions do not
correspond to \lq\lq geometrodynamical states" with nondegenerate
metric. This is because the volume operators\cite{rove}\cite{loll2}
have vanishing eigenvalues on these solutions. In order to
construct geometrodynamical solutions with nonvanishing volume,
we have to consider the lattice-loop states which have
three dimensional vertex of at least four-valent
on almost every site. This is a highly nontrivial task and
left to the future investigation. We saw, however,
that the the action of the Hamiltonian constraint on a
multi-plaquette solution completely cancels only when we consider
an infinite number of terms.
This seems to be the origin of non-normalizability of the
multi-plaquette solutions w.r.t. the induced heat-kernel measure.
We anticipate that this \lq\lq cancelation of the action
of the Hamiltonian constraint by an infinite number of terms" is a
common feature of the nontrivial solutions involving
geometrodynamical solutions. We therefore conjecture that
the geometrodynamical solutions are, if any, not normalizable
w.r.t. the induced heat-kernel measure on
$SL(2,{\bf C})$ while they may be
normalizable w.r.t. the induced Haar measure on $SU(2)$.
But this should not be taken so seriously, because
there still remains a gauge degree of freedom generated by
the Hamiltonian constraint and because
the induced heat kernel measure
is thought to be a kind of \lq\lq kinematical inner product"
which does not take account of this guage symmetry\footnote{
I am grateful to Prof. L. Smolin for pointing out this and the
following statements.}.
It is often the case that the physical states are not
normalisable w.r.t. these kinematical inner products.
The problem is then to find a genuine physical inner product
which makes the physical states normalizable and which
implements the reality conditions\cite{ashte}.

\vskip1.5cm

\noindent Acknowledgments

I would like to thank Prof. K. Kikkawa, Prof. H. Itoyama and H.
Kunitomo for helpful discussions and
careful readings of the manuscript.



\begin{thebibliography}{9}
\bibitem{ADM}
R Arnowitt, S. Deser and C. W. Misner,
in {\it \lq\lq Gravitation, An
Introduction to Current Research"},
ed. by L. Witten (John Willey and Sons, 1962) Chap. 7
\bibitem{dewitt}
B. S. Dewitt, Phys. Lev. 160 (1967) 1113
\bibitem{ashte}
A. Ashtekar,  Phys. Rev. Lett. 57 (1986) 2244 ;
Phys. Rev. D36 (1987) 295
\bibitem{jacob}
T. Jacobson and L. Smolin, Nucl. Phys. B299 (1988) 295
\bibitem{H}
V. Husain, Nucl. Phys. B313 (1989) 711
\bibitem{BP}
B. Br\"{u}gmann and J. Pullin, Nucl. Phys. B363 (1991) 221
\bibitem{ezawa}
K. Ezawa, OU-HET/217, gr-qc/9506043,
to appear in Nucl. Phys. B
\bibitem{migdal}
S. Mandelstam, Phys. Rev. 175 (1968) 1580;\\
Yu. M. Makeenko and A. A. Migdal, Nucl. Phys. B188 (1981) 269
\bibitem{loll}
R. Loll, Nucl. Phys. B444 (1995) 619
\bibitem{penrose}
R. Penrose, in {\it \lq\lq Quantum Theory and Beyond"},
ed. by. T. Bastin
(Cambridge Univ. Press, Cambridge, 1971)
\bibitem{baez}
J. C. Baez, \lq\lq Spin networks in Nonperturbative
Quantum Gravity",
gr-qc/9504036; \lq\lq Spin Network States
in Gauge Theory", gr-qc/9411007
\bibitem{RS}
C. Rovelli and L. Smolin,
\lq\lq Spin networks and quantum gravity", gr-qc/9505006,
to appear in Phys. Rev. D
\bibitem{lewa}
A. Ashtekar, J. Lewandowsli, D. Marolf, J. Mour$\tilde{\rm a}$o
and T. Thiemann, \lq\lq Quantization
of diffeomorphism invariant theories
of connections with local degrees of freedom" gr-qc/9504018
\bibitem{FK}
W. Furmanski and A. Kolawa, Nucl. Phys. B291 (1987) 594;\\
J. Kogut and L. Susskind, Phys. Rev. D11 (1975) 395
\bibitem{asle}
A. Ashtekar and J. Lewandowski, J. Math. Phys. 36 (1995) 2170
\bibitem{ash2}
A. Ashtekar, J. Lewandowski, D. Marolf, J. Mour$\tilde{\rm a}$o
and T. Thiemann, \lq\lq Coherent State transform for Spaces of
Connections", gr-qc/9412014
\bibitem{hall}
B. C. Hall, J. Func. Anal. 122 (1994) 103
\bibitem{rove}
C. Rovelli and L. Smolin, Nucl. Phys. B442 (1995) 593;\\
A. Ashtekar, C. Rovelli and L. Smolin,
Phys. Rev. Lett. 69 (1992) 446
\bibitem{loll2}
R. Loll, \lq\lq the Volume Operator in Discretized Quantum Gravity",
DFF 228/05/95, gr-qc/9506014

\end{thebibliography}
\end{document}